\documentclass[conference]{IEEEtran}
\IEEEoverridecommandlockouts
\usepackage{cite}
\usepackage{academicons}
\usepackage{hyperref}
\usepackage{amsmath,amssymb,amsfonts}
\usepackage{algorithmic}
\usepackage{graphicx}
\graphicspath{ {./images/} }
\usepackage{textcomp}
\usepackage{xcolor}
\def\BibTeX{{\rm B\kern-.05em{\sc i\kern-.025em b}\kern-.08em
    T\kern-.1667em\lower.7ex\hbox{E}\kern-.125emX}}

\newcommand{\orcid}[1]{\href{https://orcid.org/#1}{\textcolor[HTML]{A6CE39}{\aiOrcid}}}

\begin{document}

\title{Epilepsy and It's Driving Forces: Understanding the Significance Behind Epileptical Pathogenesis}

\author{\IEEEauthorblockN{1\textsuperscript{st} Shreya Shah}
\IEEEauthorblockA{\textit{dept. Applied Sciences \& Technology} \\
\textit{Gujarat Technological University (GTU)}\\
Ahmadabad, India\\
studentshreya.rcis@gmail.com}
\and
\IEEEauthorblockN{2\textsuperscript{nd} Manan Shah}
\IEEEauthorblockA{\textit{dept. of Physics} \\
\textit{Indian Institute Of Technology (IITGN)}\\
Gandhinagar, India \\
https://orcid.org/0009-0003-7996-4768}
\and
\IEEEauthorblockN{3\textsuperscript{rd} Bhavin Parekh}
\IEEEauthorblockA{\textit{dept. Applied Sciences \& Technology} \\
\textit{Gujarat Technological University (GTU)}\\
Ahmedabad, India\\
osd\_bhavin@gtu.edu.in}
}

\maketitle

\begin{abstract}
Epilepsy is a neurological disorder characterized by seizures and epileptic events intertwined with religious and personal beliefs since prehistoric times. This review paper explores the historical context and challenges in defining epilepsy, the formal definition established in 2014, and the multifaceted causes of this neurological disorder. It aims to pave the way for personalized therapeutic strategies, research advancements, and informed public health planning to enhance the lives of those affected by this complex neurological condition. In addition, this review paper focuses on the mechanisms and etiologies of epileptogenesis, categorizing them by mechanisms and the underlying causes of the disorder. The review paper provides a brief overview of the current state of the art in the diagnosis, diagnosis, treatment, and treatment of epileptiform seizures.
\end{abstract}


\section{Introduction}
Epilepsy represents a significant neurological disorder that exerts its impact globally, with India alone accounting for an estimated population of more than 12 million individuals diagnosed with epilepsy (PWE). The etiological factors that contribute to more than 70\% active epilepsy instances globally remain unidentified. This highlights an urgent need for a deeper understanding of its fundamental causes to improve diagnosis, therapeutic interventions, and possibly preventive measures\cite{b1}. This comprehensive review aims to dive into the historical evolution of epilepsy conceptualization, the challenges encountered in its definition, and the formalized delineation introduced in 2014. In addition, it examines the array of causative factors associated with this intricate neurological condition. By elucidating the enigmatic origins of epilepsy, this analysis aspires to foster the development of customized treatment methodologies, propelling forward research innovation, and facilitating strategic public health initiatives designed to elevate the quality of life for individuals managing epilepsy. Historically, the precise definition of epilepsy posed significant challenges, as seizures and epileptic occurrences were often enmeshed with religious and individual beliefs tracing back to prehistoric eras. The absence of a definitive and unambiguous definition proved to be an impediment to the effective study and management of this medical concern.\\
\\
In 2014, a comprehensive and formal definition was formulated, designating epilepsy as a neurological disorder of the brain. This definition specifies several criteria: the occurrence of a minimum of two spontaneous (or reflex) seizures separated by an interval exceeding 24 hours; at least one spontaneous (or reflex) seizure coupled with an estimated recurrence risk of no less than 60\% after having two unprovoked seizures within a decade; or the identification of an Epilepsy syndrome\cite{b2}. Epileptic syndromes are typified by a spectrum of characteristics, which encompass various forms of seizures, findings from electroencephalograms (EEGs), and imaging studies. These syndromes frequently exhibit traits influenced by the patient's age and are associated with distinct co-morbid conditions\cite{b3}. A thorough comprehension of the multitude of factors responsible for epilepsy is essential for accurate diagnosis, the customization of therapeutic interventions, and potential preventative strategies. This scholarly review aspires to enhance the management and prognosis of epilepsy patients, offering optimism for a more promising future in the realm of epilepsy treatment.

\section{Mechanisms of Epileptogenesis}

\subsection{Excitation-Inhibition Imbalance} 
Epileptogenesis involves alterations in the balance between excitatory and inhibitory signals, leading to hyperexcitability and an increased risk of seizures. Changes in ion channels, neurotransmitter systems, and neural circuit connectivity contribute to this imbalance, making the brain more susceptible to seizures.

\subsection{Role of Ion Channels} 
Ion channels are critical regulators of neuronal excitability and play a significant role in epilepsy development. Dysregulation or mutations in ion channels can lead to abnormal electrical signaling, promoting seizure generation. Understanding the role of ion channels is vital for identifying potential therapeutic targets and developing more effective antiepileptic drugs.

\subsection{Neuroinflammation and Epilepsy} 
Inflammation in the brain, known as neuroinflammation, is emerging as a crucial player in epilepsy pathogenesis. Brain injuries, infections, or autoimmune reactions can trigger inflammatory responses, disrupting the balance between excitation and inhibition and contributing to seizure development. Targeting neuroinflammatory pathways may open new avenues for antiepileptic therapies and disease modification. In conclusion, understanding the intricate mechanisms of epileptogenesis, including excitatory-inhibitory imbalance, ion channel dysregulation, and neuroinflammation, is crucial for developing targeted therapies to control seizures and improve the quality of life for individuals with epilepsy. Ongoing research offers hope for innovative treatments and better epilepsy management in the future.

\section{Mechanistic Causes}
\begin{figure}[h]
    \centering
    \includegraphics[width=0.5\textwidth]{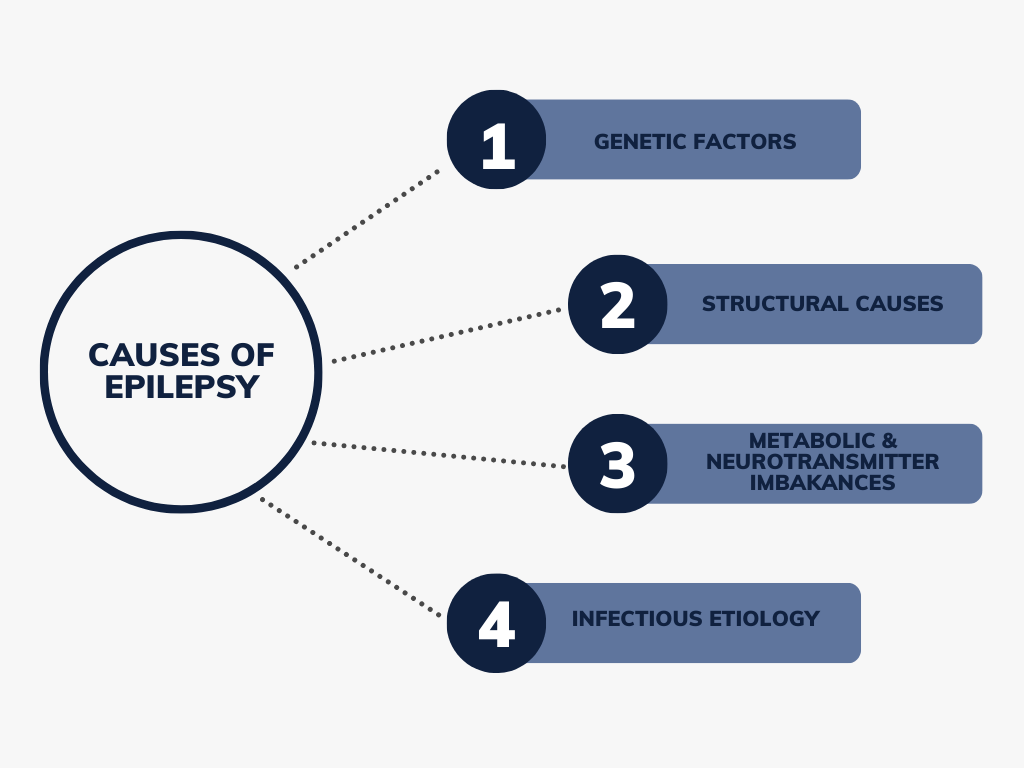}
    \caption{A Comprehensive Framework Including Genetic, Structural, Metabolic/Neurotransmitter, and Infectious Cause}
    \label{fig:mesh1}
\end{figure}

\subsection{Genetic Factors}
The field of epilepsy research has made significant progress in understanding the genetic basis of the disorder, leading to the emergence of precision medicine as a potential approach for tailored therapeutic interventions. Precision medicine involves the classification of epilepsy patients into distinct subpopulations based on genetic differences, allowing for targeted treatments that minimize side effects and unnecessary costs\cite{b4}. Genetic diagnoses play a pivotal role in identifying treatment-relevant patient subgroups, especially in cases where specific pathogenic variants can be targeted for intervention. Additionally, research on noncoding regions of the genome has provided valuable insights into disease development, contributing to our understanding of various epilepsy syndromes. Efforts by international research teams, such as Epi4K, the Epilepsy Phenome/Genome Project, and EuroEPINOMICS, have greatly accelerated the identification of genes associated with epilepsy\cite{b5}\cite{b6}. Advances in bioinformatics tools like MME and DECIPHER have further facilitated the rapid growth of the known epilepsy-associated genes\cite{b7}. Moreover, discoveries related to noncoding regions of the genome have shed light on the genetic mechanisms underlying specific epilepsy disorders. Variants outside of annotated coding regions, like those identified in the SCN1A gene associated with Dravet syndrome, have been found to influence gene expression and disease development\cite{b8}. In addition to the well-studied Dravet syndrome, research has elucidated the genetic basis of other epilepsy syndromes, such as Adult Familial Myoclonic Epilepsy, through the identification of intronic expansions in the SAMD12 gene. These findings have significant implications for targeted therapies and prognostication, bringing us closer to achieving precision medicine's potential in epilepsy management\cite{b9}.\\
\\
The increasing understanding of the genetic roots of epilepsy has initiated a transformational era in precision medicine aimed at this neurological condition.  The capacity to classify individuals into distinct subgroups based on genetic variations has facilitated the creation of personalised therapy strategies, therefore increasing treatment effectiveness and increasing overall quality of life for affected persons. International collaborations, along with advances in bioinformatics, have significantly made the finding of genes associated with epilepsy easier, while studies of the noncoding areas of the genome have revealed profound insights into the mechanisms of the disease.  As our understanding of the genetic foundations of epilepsy grows, precision medicine emerges as a key advancement in epilepsy therapy, bringing us closer to the goal of better seizure control and general well-being for afflicted individuals.
\subsection{Structural Causes}
Structural causes of epilepsy encompass various conditions, including developmental disorders of the brain during fetal or childhood growth, brain tumours, and mesial temporal sclerosis. Cortical dysplasia, a result of irregular neuron migration during fetal development, can lead to abnormal communication and recurring seizures. A rare example is hemimegalencephaly, characterized by one hemisphere of the brain being larger than the other, causing frequent seizures and developmental delays\cite{b10}. Treatment may involve anti-seizure medications and, if necessary, surgical removal of the affected hemisphere, showcasing the brain's remarkable neuroplasticity. Patients with brain tumours often experience epilepsy as a comorbid condition, with tumour-related seizures being less responsive to traditional antiepileptic medications. Antineoplastic therapies, such as surgical resection, radiation, and chemotherapy, are being increasingly recognized for their role in both tumour management and epilepsy control\cite{b11}. Mesial temporal sclerosis, triggered by head trauma, brain infection, stroke, or aneurysm, can lead to scar formation in the temporal lobe, causing temporal lobe epilepsy with partial seizures. Treatment options include anti-seizure medications, dietary adjustments, surgery, or nerve stimulation to manage associated epilepsy and improve patient well-being\cite{b12}.\\
\\
Structural causes of epilepsy present complex challenges, requiring a multifaceted approach for diagnosis and treatment. Advancements in understanding developmental disorders, brain tumors, and mesial temporal sclerosis have shed light on the mechanisms underlying epilepsy in these conditions. Tailored therapeutic strategies, such as surgical interventions and antineoplastic therapies, hold promise in managing epilepsy alongside the primary structural conditions. As research progresses, innovative treatments and targeted therapies may further enhance epilepsy control and improve the quality of life for patients affected by these structural causes of epilepsy. Continued collaborative efforts between neurologists, oncologists, and researchers are crucial in advancing knowledge and optimizing care for individuals living with epilepsy in the context of these structural brain conditions.

\subsection{Metabolic \& Neurotransmitter Imbalances}
Metabolic and neurotransmitter imbalance causes of epilepsy encompass various disorders, including mitochondrial dysfunction, amino acid metabolism disorders, and disturbances in glutamatergic transmission. Mitochondria play a crucial role in ATP production and several other cellular functions, and their dysfunction has been linked to seizure generation through changes in calcium homeostasis, oxidative effects on ion channels and neurotransmitter transporters, decreased neuronal membrane potential, and impaired network inhibition\cite{b13}. Additionally, disorders of amino acid metabolism, such as classical maple syrup urine disease, may lead to seizures, highlighting the importance of maintaining appropriate amino acid levels\cite{b14}. Emphasis on proper regulation of amino acid levels such as that of glutamate and GABA must also be highlighted since they act as neurotransmitters and as an energy source for the brain. In recent years, substantial progress has been made in comprehending the role of excitatory glutamatergic transmission in seizures. Glutamate has emerged as a key participant in the initiation, propagation, and maintenance of epileptic activity. Epileptic patients exhibit alterations in glutamate concentration and receptor function. Studies have demonstrated that glutamate antagonists display strong anticonvulsant properties in diverse experimental seizure models, including those of genetic, chemical, electrical, or kindling-induced origins. Although the precise therapeutic implications of drugs targeting glutamatergic mechanisms in epilepsy are yet to be fully defined, the potential for such treatments continues to pique interest and warrants further investigation\cite{b15}.\\
\\
Understanding the underlying metabolic and neurotransmitter imbalances contributing to epilepsy is crucial for developing targeted treatments and improving patient outcomes. Advances in the study of mitochondrial disorders, amino acid metabolism, and glutamatergic transmission have shed light on the intricate relationships between these factors and epileptic activity. As research continues, exploring the potential of therapeutic interventions that address these imbalances may lead to novel and more effective approaches for managing epilepsy and providing relief to those affected by this complex neurological condition. Collaborative efforts between clinicians, researchers, and medical experts are essential in unraveling the intricacies of metabolic and neurotransmitter causes of epilepsy, ultimately striving towards better care and improved quality of life for individuals living with epilepsy.
\subsection{Infectious Etiology}

Central nervous system (CNS) infections pose a significant risk factor for epilepsy, with unprovoked seizure rates ranging from 6.8\% to 8.3\% in developed countries and higher prevalence in resource-poor regions. Infections and infestations are among the most prevalent preventable risk factors for seizures and acquired epilepsy worldwide. Approximately 60 million of the 70 million people living with epilepsy reside in low- to medium-income countries, highlighting the urgent need for comprehensive strategies to address this burden\cite{b16}. A wide array of CNS infections, including bacterial (e.g. bacterial meningitis, tuberculosis), viral (e.g., herpes simplex, HHV-6), parasitic (e.g., cerebral toxoplasmosis, NCC, malaria), fungal (e.g. candidiasis, coccidioidomycosis, aspergillosis), and prion infections (CJD), can trigger status epilepticus, a severe and prolonged epileptic seizure activity. The mechanisms of a viral infection like that of the Herpes Simplex Virus or that with HHV-6, involve viral replication leading to inflammation and immunopathology, contributing to neuronal damage or dysfunction. Understanding the role of viral infections in the CNS and their association with epilepsy is crucial for the diagnosis, prevention, and targeted treatment of seizure disorders\cite{b17}.\\
\\
Bacterial infections primarily affecting the meninges and cerebral parenchyma are another significant risk factor for epilepsy. Acute symptomatic seizures are well known in survivors of acute bacterial meningitis, and efforts to characterize the risk of seizures based on the infective agent are essential\cite{b18}. Preventive measures, such as vaccination programs against the major meningeal pathogens, can significantly reduce the burden of bacterial meningitis and mitigate the associated epilepsy risk. Timely diagnosis and effective antibiotic therapy, coupled with adjuvant therapies to regulate mediators involved in neuronal injury, can further reduce the likelihood of subsequent epilepsy development in survivors of bacterial meningitis\cite{b19}.\\
\\
Additionally, parasitosis conditions are linked to seizures and acquired epilepsy, with cysticercosis being a common neurological infestation. Cysticercosis, caused by the larval form of Taenia solium with pigs as the intermediate host, is a prevalent neurological infestation and a significant risk factor for acquired epilepsy in numerous African, Asian, and Latin American countries\cite{b20}. Understanding the mechanisms of epileptogenesis in cysticercosis, involving factors like inflammation, oedema, gliosis, and genetic interactions, is critical for improved diagnosis and management of epilepsy in affected individuals\cite{b21}. Addressing infectious etiologies of epilepsy globally requires comprehensive strategies, including preventive measures, early diagnosis, and effective treatments, particularly in resource-poor settings. Collaborative efforts and research advancements in this domain are pivotal to alleviating the impact of infections on epilepsy, ultimately enhancing the well-being and quality of life for those affected by this neurological disorder.

\section{Epilepsies Etiologically}
\subsection{Acquired Epilepsies}
\begin{figure}[h]
    \centering
    \includegraphics[width=0.5\textwidth]{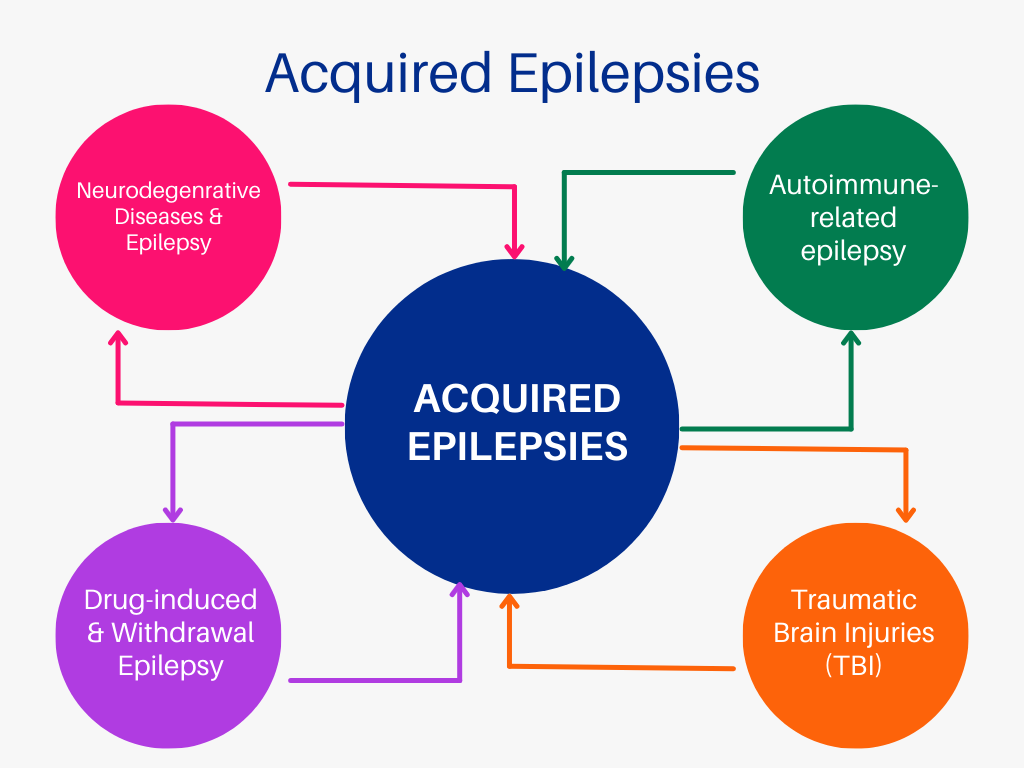}
    \caption{Classification of Acquired Epilepsies: Four Major Categories Including Neurodegenerative, Autoimmune, Traumatic, and Drug-induced Condition}
    \label{fig:mesh2}
\end{figure}
\begin{subsubsection}{Autoimmune-Related Epilepsy}
    A notable subset of patients with epilepsy of unknown origin has been linked to autoimmune causes. Recent breakthroughs in research have illuminated the role of autoimmunity in epilepsy, leading to its recognition as a distinct entity in the 2017 epilepsy classification by the International League Against Epilepsy (ILAE) \cite{b22}. Neural-specific antibodies have emerged as pivotal players in understanding the etiopathogenesis of autoimmune epilepsy, with some cases associated with occult tumors expressing onconeural antigens, triggering misdirected immune responses that contribute to neuronal dysfunction. Specific antibody responses, such as ANNA-1 IgG in small cell lung carcinoma and Ma-2 IgG in testicular germ cell cancer, are tumor-specific, providing valuable insights into the complex interplay between autoimmunity and epilepsy \cite{b23}.
    Infections have also been identified as potential triggers for autoimmune neurological syndromes, with molecular mimicry, epitope spreading, and bystander activation among the proposed mechanisms \cite{b24}. In some cases, infections may lead to cellular damage and the release of autoantigens, prompting an autoimmune response. Additionally, infections can activate autoreactive lymphocytes and antigen-presenting cells, further contributing to an immune response against self-antigens. A well-known example is the association between herpes simplex virus encephalitis and NMDA-R encephalitis, highlighting the significance of recognizing infectious triggers for autoimmune epilepsy and their potential implications for prevention and treatment \cite{b25}.
    Autoantibodies play a significant role in autoimmune-related epilepsies, disrupting normal neuronal function and contributing to hyperexcitability and seizure development. Understanding the pathogenic mechanisms underlying autoimmunity in epilepsy is crucial for accurate diagnosis and the development of targeted therapies to manage seizure activity and improve patient outcomes. The recognition of specific autoantibodies and potential infectious triggers offers promising opportunities to advance the diagnosis and treatment of autoimmune-related epilepsies, potentially transforming the landscape of epilepsy care for affected individuals.
\end{subsubsection}
\\
\subsubsection{Traumatic Brain Injury}
Posttraumatic epilepsy (PTE) is a significant neurodegenerative condition responsible for approximately 20\% of symptomatic epilepsy cases. Individuals who have experienced head injuries due to falls, car accidents, sports-related incidents, or other accidents are at a higher risk of developing seizures or epilepsy. The likelihood of developing this condition increases with the number of head traumas, and genetic factors may also play a role in its onset. PTE is characterized by a complex pathophysiology involving structural brain changes, scar formation, neuronal damage, and inflammatory responses. Despite its debilitating impact, there is hope for early intervention as PTE often exhibits a long latent phase before seizures manifest, allowing for potential prophylactic treatment strategies if high-risk patients can be identified\cite{b26}.

Researchers have identified some promising imaging biomarker candidates for predicting posttraumatic epileptogenesis, but further investigation is required to ensure their specificity and sensitivity for accurate prediction. The ability to detect early signs of epileptogenesis could provide a critical window for preventive interventions to delay or halt the onset of epilepsy\cite{b27}. In addition to traditional treatment approaches like antiepileptic medications, ketogenic diets, surgery, and neurostimulation, efforts to identify individuals at risk for PTE before seizure manifestation could significantly improve patient outcomes and enhance the overall quality of life for those affected.

Continued research into the underlying mechanisms of posttraumatic epileptogenesis, combined with effective head injury prevention measures, will be instrumental in reducing the incidence and burden of PTE. The development of reliable imaging biomarkers and a better understanding of the complex genetic factors involved will pave the way for personalized treatment strategies and improved patient care. Early diagnosis and timely interventions offer hope for mitigating the impact of posttraumatic epilepsy, providing a brighter outlook for affected individuals and their families.
\\
\subsubsection{Drug-Induced \& Withdrawal Epilepsy}
Patients diagnosed with epilepsy often face an increased risk of experiencing psychiatric co-morbidities, including psychotic disorders, which can further burden their overall health. Research on a Danish population-based cohort revealed significantly higher incidences of schizophrenia and schizophrenia-like psychosis in epilepsy patients compared to the general population\cite{b28}. Additionally, antiepileptic drug-induced psychotic disorder (AIPD) poses challenges, with reported prevalence ranging from 1.0\% to 8.4\% in clinical trials of antiepileptic drugs. Despite its significance, standardized definitions or diagnostic criteria for AIPD are currently lacking, highlighting the need for better collaboration between epileptology and psychiatry to establish clearer approaches for diagnosis and treatment\cite{b29}.

Seizures frequently manifest in individuals with substance abuse issues, and their occurrence can be attributed to various mechanisms, either indirectly through factors like CNS infections, cerebral trauma, stroke, or metabolic imbalances, or directly due to intoxication or withdrawal from substances. These mechanisms are not mutually exclusive, warranting a comprehensive approach to management. Identifying medical and surgical emergencies is crucial, along with addressing nonconvulsive signs of intoxication and withdrawal. For drug withdrawal, using agents from the same pharmacological class or with cross-tolerance is recommended. Long-term anticonvulsant prophylaxis is generally not indicated when seizures are solely attributed to drug intoxication or withdrawal. Tailored approaches are essential to optimize patient outcomes and effectively manage seizures in substance abusers\cite{b30}.

In conclusion, psychiatric comorbidities and AIPD present significant challenges in epilepsy management, necessitating multidisciplinary efforts to improve understanding, diagnosis and treatment. Collaborative research between epileptology and psychiatry is crucial to establish standardized criteria for AIPD and improve patient care. In addition, a comprehensive approach is essential in the management of seizures in substance abusers, considering the various contributing factors and optimising treatment strategies for better patient outcomes. By addressing these complexities, healthcare professionals can enhance the quality of life of individuals with epilepsy and reduce the burden of psychiatric comorbidities.

\subsubsection{Neurodegenerative Diseases \& Epilepsy}
Both epilepsy and Alzheimer's disease (AD) are common neurological disorders that become more prevalent with age. Evidence suggests an intriguing interaction between these two conditions, as patients with AD face an increased risk of developing seizures and epilepsy. Neurodegenerative conditions, including AD, are presumed to be the etiology for a significant proportion of epilepsy cases in older populations. Moreover, neurodegenerative disorders are responsible for approximately 10\% of new-onset epilepsy cases in patients older than 65.
Pathological Changes in Mesial Temporal Structures: In both AD and temporal lobe epilepsy, pathological alterations occur in mesial temporal structures, specifically the CA1, subiculum, and entorhinal cortex. In temporal lobe epilepsy models, these changes result in cell loss and reorganization of neuronal circuitry, contributing to pathological hyperexcitability in the affected regions. In addition, patients with AD and other dementias may exhibit substantial cell loss and gliosis in the CA1 region\cite{b31}. 
Temporal lobe epilepsy and Parkinson's disease are characterized by disturbances in gamma-amino butyric acid (GABA) signaling, the primary inhibitory neurotransmitter in the central nervous system (CNS). In temporal lobe epilepsy, seizures reflect excessive excitation due to potential dysfunction in local inhibitory circuits. Conversely, Parkinson's disease disrupts the input to striatal GABAergic neurons, leading to significant impairments in GABA signaling within the CNS \cite{b32}.
Both Alzheimer's disease and Parkinson's disease are neurodegenerative disorders associated with the loss of locus coeruleus (LC) noradrenergic neurons. These disorders share two hallmark symptoms: cognitive impairment and depression. The hippocampus, crucial for cognition and mood regulation, receives exclusive noradrenergic innervation from LC neurons. However, the exact mechanisms by which the loss of LC noradrenergic neurons contributes to these shared symptoms in AD and PD remain to be fully elucidated\cite{b33}.
Norepinephrine, a prominent neurotransmitter in the CNS, is widely distributed across numerous brain regions. The locus coeruleus houses over half of all noradrenergic neurons, projecting throughout the CNS via five major tracts. Different classes and subtypes of adrenoceptors, coupled with diverse G proteins, participate in the complex noradrenergic signaling system. The hippocampus and cortex receive sole noradrenergic innervation from the LC, exerting various effects depending on the adrenoreceptor targeted\cite{b34}.
An altered noradrenergic nervous system is known to be involved in psychiatric disorders (depression, attention deficit disorder, Tourette's syndrome, and psychosis), neurologic conditions (epilepsy, Parkinson’s disease, and Alzheimer’s disease), posttraumatic stress disorder (PTSD), and sleep disorders\cite{b35}. Understanding the intricate interactions between neurodegenerative disorders, epilepsy, and the noradrenergic nervous system presents a fascinating area of research. Further investigations into the complex connections and signaling pathways hold promise for advancing our comprehension of these neurological and psychiatric conditions and potentially unveiling new therapeutic strategies.

\subsection{Idiopathic and Cryptogenic Epilepsy}
Diagnosing epilepsy can be intricate, especially when the cause is not apparent, as seen in idiopathic and cryptogenic epilepsies. The absence of clear brain abnormalities or identifiable lesions requires reliance on clinical history, neurological examinations, and EEG findings for diagnosis. However, the diverse clinical presentations and lack of visible abnormalities can complicate the distinction between idiopathic and cryptogenic epilepsy\cite{b36}. In patients over 65 years old, neurodegenerative conditions account for approximately 10\% of new-onset epilepsy cases.

To address these diagnostic challenges, researchers and clinicians are actively engaged in ongoing efforts to better understand idiopathic and cryptogenic epilepsies. Advances in genetic research have identified various genetic mutations contributing to idiopathic epilepsies, facilitating targeted therapies. For cryptogenic epilepsies, research focuses on unraveling unknown causes, exploring subtle brain abnormalities, and disruptions in molecular pathways using advanced neuroimaging techniques like fMRI and PET. Collaborative interdisciplinary efforts, integrating clinical, genetic, and neuroimaging data, offer a comprehensive understanding of the underlying mechanisms.

In conclusion, diagnosing idiopathic and cryptogenic epilepsies requires meticulous evaluation, and advances in genetic and neuroimaging technologies have opened new avenues for research. A better understanding of the underlying mechanisms holds promise for personalized and targeted treatments, improving outcomes for individuals affected by these complex epileptic disorders. Collaborative interdisciplinary approaches play a crucial role in overcoming diagnostic challenges and guiding future therapeutic strategies for these conditions.

\section{Towards Personalized Therapeutic Strategies}
Personalized therapeutic strategies in epilepsy aim to optimize treatment approaches based on individual patient characteristics, seizure etiology, and treatment responses.

\subsection{Targeting Underlying Causes}
Personalized therapeutic strategies in epilepsy involve identifying the specific cause of epilepsy in each patient to guide treatment decisions. For genetic epilepsies, targeted therapies addressing specific gene mutations may be more effective. In cases related to brain tumors or vascular malformations, surgical intervention may be preferred. Understanding underlying causes enables tailored treatment, leading to better seizure control and improved quality of life.

\subsection{Precision Medicine Approaches}
Precision medicine uses genetic testing and biomarkers to individualize epilepsy treatments. Genetic testing identifies gene variants influencing drug responses, guiding the choice of antiepileptic drugs. Pharmacogenomics reduces trial-and-error, leading to more successful seizure management. Precision medicine may also identify patients benefiting from alternative therapies like ketogenic diets or neuromodulation techniques.

\subsection{Epilepsy Biomarkers}
Research in epilepsy biomarkers is ongoing to provide valuable insights for diagnosis, prognosis, and treatment response. Neuroimaging, EEG, and blood-based assays show promise in identifying potential biomarkers. Specific brain activity patterns observed through fMRI or EEG can differentiate seizure types. Blood-based biomarkers, such as certain proteins or genetic markers, may indicate disease progression or treatment response. Reliable epilepsy biomarkers could revolutionize management, allowing early diagnosis and predicting patient outcomes.

In conclusion, personalized therapeutic strategies in epilepsy offer patient-centered approaches for optimized seizure management by identifying underlying causes, employing precision medicine, and exploring epilepsy biomarkers. These advancements hold promise for transforming epilepsy care and enhancing the lives of individuals with epilepsy.

\section{Conclusion \& Further Research}
Epilepsy is acknowledged as a multifaceted neurological condition with numerous etiologies, encompassing genetic mutations, structural brain anomalies, and disruptions in neurotransmitter function. The diagnosis of epilepsy proves particularly challenging in instances where the underlying cause is not readily discernible, stimulating continued scientific inquiry into idiopathic and cryptogenic forms of epilepsy. Strategies that cater to individual patient needs, such as precision medicine and the identification of specific biomarkers, show great promise in refining the management of epilepsy and improving seizure control.,Further investigation is imperative to fully grasp the complexity and consequences stemming from the factors that drive epilepsy. Comprehensive analyses in the fields of genetics, advanced neuroimaging techniques, and biomarker research have the potential to pinpoint distinct subtypes of epilepsy, facilitating personalized therapeutic interventions that aim to enhance patient prognoses.,Currently, the absence of a definitive preventative measure or cure for epilepsy persists, yet innovative pharmacological agents targeting the process of epileptogenesis, including compounds like rapamycin, neurosteroids, and specific pathway inhibitors, exhibit significant potential. Nonetheless, contemporary research underscores the necessity for more efficacious treatment modalities, with exploration into pathophysiology-based strategies being pursued to discover novel drugs capable of curing epilepsy and enhancing the quality of life for those affected by the disorder \cite{b37}.



\begin{thebibliography}{00}
\bibitem{b1}Amudhan, S., Gururaj, G., \& Satishchandra, P. (2015). Epilepsy in India I: Epidemiology and public health. Annals of Indian Academy of Neurology, 18(3), 263–277. https://doi.org/10.4103/0972-2327.160093
\bibitem{b2} Fisher, R. S., Acevedo, C., Arzimanoglou, A., Bogacz, A., Cross, J. H., Elger, C. E., ... \& Wiebe, S. (2014). ILAE official report: a practical clinical definition of epilepsy. Epilepsia, 55(4), 475-482.
\bibitem{b3} Wirrell, E. C., Nabbout, R., Scheffer, I. E., Alsaadi, T., Bogacz, A., French, J. A., ... \& Tinuper, P. (2022). Methodology for classification and definition of epilepsy syndromes with list of syndromes: report of the ILAE Task Force on Nosology and Definitions. Epilepsia, 63(6), 1333-1348.
\bibitem{b4} National Research Council. (2011). Toward precision medicine: building a knowledge network for biomedical research and a new taxonomy of disease.
\bibitem{b5} Sherr, E., Lowenstein, D., Kirsch, H., Alldredge, B., Allen, A. S., Berkovic, S. F., ... \& Han, Y. (2013). De novo mutations in epileptic encephalopathies.
\bibitem{b6} Allen, A. S., Bellows, S. T., Berkovic, S. F., Bridgers, J., Burgess, R., Cavalleri, G., ... \& Winawer, M. R. (2017). Ultra-rare genetic variation in common epilepsies: a case-control sequencing study. The Lancet Neurology, 16(2), 135-143.
\bibitem{b7} Azzariti, D. R., \& Hamosh, A. (2020). Genomic data sharing for novel mendelian disease gene discovery: the matchmaker exchange. Annual Review of Genomics and Human Genetics, 21, 305-326.
\bibitem{b8} Carvill, G. L., Engel, K. L., Ramamurthy, A., Cochran, J. N., Roovers, J., Stamberger, H., ... \& Mefford, H. C. (2018). Aberrant inclusion of a poison exon causes dravet syndrome and related SCN1A-associated genetic epilepsies. The American Journal of Human Genetics, 103(6), 1022-1029.
\bibitem{b9} Ishiura, H., Doi, K., Mitsui, J., Yoshimura, J., Matsukawa, M. K., Fujiyama, A., ... \& Tsuji, S. (2018). Expansions of intronic TTTCA and TTTTA repeats in benign adult familial myoclonic epilepsy. Nature genetics, 50(4), 581-590.
\bibitem{b10} Baldassari, S., Ribierre, T., Marsan, E., Adle-Biassette, H., Ferrand-Sorbets, S., Bulteau, C., Dorison, N., Fohlen, M., Polivka, M., Weckhuysen, S., Dorfmüller, G., Chipaux, M., \& Baulac, S. (2019). Dissecting the genetic basis of focal cortical dysplasia: a large cohort study. Acta neuropathologica, 138(6), 885–900. 
\bibitem{b11} Rudà, R., Trevisan, E., \& Soffietti, R. (2010). Epilepsy and brain tumors. Current opinion in oncology, 22(6), 611-620.
\bibitem{b12} Curia, G., Lucchi, C., Vinet, J., Gualtieri, F., Marinelli, C., Torsello, A., Costantino, L., \& Biagini, G. (2014). Pathophysiogenesis of mesial temporal lobe epilepsy: is prevention of damage antiepileptogenic?. Current medicinal chemistry, 21(6), 663–688.
\bibitem{b13} Zsurka, G., \& Kunz, W. S. (2015). Mitochondrial dysfunction and seizures: the neuronal energy crisis. The Lancet Neurology, 14(9), 956-966.
\bibitem{b14} Lee, W. T. (2011). Disorders of amino acid metabolism associated with epilepsy. Brain and Development, 33(9), 745-752.
\bibitem{b15} Urbanska, E. M., Czuczwar, S. J., Kleinrok, Z., \& Turski, W. A. (1998). Excitatory amino acids in epilepsy. Restorative neurology and neuroscience, 13(1-2), 25-39.
\bibitem{b16} Singhi P. (2011). Infectious causes of seizures and epilepsy in the developing world. Developmental medicine and child neurology, 53(7), 600–609. 
\bibitem{b17} Vezzani, A., Fujinami, R. S., White, H. S., Preux, P. M., Blümcke, I., Sander, J. W., \& Löscher, W. (2016). Infections, inflammation and epilepsy. Acta neuropathologica, 131(2), 211–234.
\bibitem{b18} Aminoff, M. J., Boller, F., Bruyn, G. W., Klawans, H. L., Swaab, D. F., \& Vinken, P. J. (Eds.). (1968). Handbook of clinical neurology. North-Holland Publishing Company.
\bibitem{b19} Murthy, J. M. K., \& Prabhakar, S. (2008). Bacterial meningitis and epilepsy. Epilepsia, 49, 8-12.
\bibitem{b20} Bruno, E., Bartoloni, A., Zammarchi, L., Strohmeyer, M., Bartalesi, F., Bustos, J. A., Santivañez, S., García, H. H., Nicoletti, A., \& COHEMI Project Study Group (2013). Epilepsy and neurocysticercosis in Latin America: a systematic review and meta-analysis. PLoS neglected tropical diseases, 7(10), e2480. https://doi.org/10.1371/journal.pntd.0002480
\bibitem{b21} Nash, T. E., Mahanty, S., Loeb, J. A., Theodore, W. H., Friedman, A., Sander, J. W., Singh, G., Cavalheiro, E., Del Brutto, O. H., Takayanagui, O. M., Fleury, A., Verastegui, M., Preux, P. M., Montano, S., Pretell, E. J., White, A. C., Jr, Gonzales, A. E., Gilman, R. H., \& Garcia, H. H. (2015). Neurocysticercosis: A natural human model of epileptogenesis. Epilepsia, 56(2), 177–183. 
\bibitem{b22} Scheffer, I. E., Berkovic, S., Capovilla, G., Connolly, M. B., French, J., Guilhoto, L., Hirsch, E., Jain, S., Mathern, G. W., Moshé, S. L., Nordli, D. R., Perucca, E., Tomson, T., Wiebe, S., Zhang, Y. H., \& Zuberi, S. M. (2017). ILAE classification of the epilepsies: Position paper of the ILAE Commission for Classification and Terminology. Epilepsia, 58(4), 512–521.
\bibitem{b23} Husari, K. S., \& Dubey, D. (2019). Autoimmune epilepsy. Neurotherapeutics, 16, 685-702.
\bibitem{b24} Fujinami, R. S., von Herrath, M. G., Christen, U., \& Whitton, J. L. (2006). Molecular mimicry, bystander activation, or viral persistence: infections and autoimmune disease. Clinical microbiology reviews, 19(1), 80–94.
\bibitem{b25} Armangue, T., Spatola, M., Vlagea, A., Mattozzi, S., Cárceles-Cordon, M., Martinez-Heras, E., Llufriu, S., Muchart, J., Erro, M. E., Abraira, L., Moris, G., Monros-Giménez, L., Corral-Corral, Í., Montejo, C., Toledo, M., Bataller, L., Secondi, G., Ariño, H., Martínez-Hernández, E., Juan, M., … Spanish Herpes Simplex Encephalitis Study Group (2018). Frequency, symptoms, risk factors, and outcomes of autoimmune encephalitis after herpes simplex encephalitis: a prospective observational study and retrospective analysis. The Lancet. Neurology, 17(9), 760–772.
\bibitem{b26} Agrawal, A., Munivenkatappa, A., Shukla, D. P., Menon, G. R., Alogolu, R., Galwankar, S., Kumar, S. S., Momhan, P. R., Pal, R., \& Rustagi, N. (2016). Traumatic brain injury related research in India: An overview of published literature. International journal of critical illness and injury science, 6(2), 65–69. 
\bibitem{b27} Immonen, R., Harris, N. G., Wright, D., Johnston, L., Manninen, E., Smith, G., ... \& Grohn, O. (2019). Imaging biomarkers of epileptogenecity after traumatic brain injury–Preclinical frontiers. Neurobiology of disease, 123, 75-85.
\bibitem{b28} Brust, J. C. (2006). Seizures and substance abuse: treatment considerations. Neurology, 67(12 suppl 4), S45-S48.
\bibitem{b29} Chen, Z., Lusicic, A., O’Brien, T. J., Velakoulis, D., Adams, S. J., \& Kwan, P. (2016). Psychotic disorders induced by antiepileptic drugs in people with epilepsy. Brain, 139(10), 2668-2678.
\bibitem{b30} Piedad, J., Rickards, H., Besag, F. M., \& Cavanna, A. E. (2012). Beneficial and adverse psychotropic effects of antiepileptic drugs in patients with epilepsy: a summary of prevalence, underlying mechanisms and data limitations. CNS drugs, 26(4), 319–335.
\bibitem{b31} Friedman, D., Honig, L. S., \& Scarmeas, N. (2012). Seizures and epilepsy in Alzheimer's disease. CNS neuroscience \& therapeutics, 18(4), 285-294.
\bibitem{b32} Kleppner, S. R., \& Tobin, A. J. (2001). GABA signalling: therapeutic targets for epilepsy, Parkinson’s disease and Huntington’s disease. Emerging Therapeutic Targets, 5(2), 219-239.
\bibitem{b33} Szot, P. (2012). Common factors among Alzheimer’s disease, Parkinson’s disease, and epilepsy: possible role of the noradrenergic nervous system. Epilepsia, 53, 61-66.
\bibitem{b34} Aston-Jones, G., Rajkowski, J., \& Cohen, J. (2000). Locus coeruleus and regulation of behavioral flexibility and attention. Progress in brain research, 126, 165–182. 
\bibitem{b35} Leckman, J. F., Hardin, M. T., Riddle, M. A., Stevenson, J., Ort, S. I., \& Cohen, D. J. (1991). Clonidine treatment of Gilles de la Tourette's syndrome. Archives of general psychiatry, 48(4), 324–328.
\bibitem{b36} Engel, J., \& Pedley, T. A. (2008). Introduction: what is epilepsy. Epilepsy: a comprehensive textbook, 2nd edition. Philadelphia: Lippincott Williams \& Wilkins, 1-13.
\bibitem{b37} Clossen, B. L., \& Reddy, D. S. (2017). Novel therapeutic approaches for disease-modification of epileptogenesis for curing epilepsy. Biochimica et Biophysica Acta (BBA)-Molecular Basis of Disease, 1863(6), 1519-1538.
\end{thebibliography}
\end{document}